\documentclass[legalpaper,twocolumn]{article}

\usepackage{amssymb,amsfonts,amsmath,amstext,amsgen,amsopn,amsxtra,indentfirst,graphicx}

\begin{document}

\title{Supervised Machine Learning for Analysing Spectra of Exoplanetary Atmospheres}
\date{}
\maketitle
\vspace{-0.5in}
\noindent
\author{Pablo M\'arquez-Neila$^{1,2}$, Chloe Fisher$^2$, \\ Raphael Sznitman$^1$, Kevin Heng$^{2,3}$}\\
\author{\scriptsize 1: University of Bern, ARTORG Center for Biomedical Engineering, Murtenstrasse 50, CH-3008, Bern, Switzerland}\\
\author{\scriptsize 2: University of Bern, Center for Space and Habitability, Gesellschaftsstrasse 6, CH-3012, Bern, Switzerland}\\
\author{\scriptsize 3: Corresponding author: kevin.heng@csh.unibe.ch}

\vspace{0.2in}

\textbf{The use of machine learning is becoming ubiquitous in astronomy \cite{b10,graff14,pearson18}, but remains rare in the study of the atmospheres of exoplanets.  Given the spectrum of an exoplanetary atmosphere, a multi-parameter space is swept through in real time to find the best-fit model \cite{madhu09,bs12,line12}.  Known as ``atmospheric retrieval", it is a technique that originates from the Earth and planetary sciences \cite{rodgers00}.  Such methods are very time-consuming and by necessity there is a compromise between physical and chemical realism versus computational feasibility.  Machine learning has previously been used to determine which molecules to include in the model, but the retrieval itself was still performed using standard methods \cite{waldmann16}.  Here, we report an adaptation of the ``random forest" method of supervised machine learning \cite{ho98,breiman01}, trained on a pre-computed grid of atmospheric models, which retrieves full posterior distributions of the abundances of molecules and the cloud opacity.  The use of a pre-computed grid allows a large part of the computational burden to be shifted offline.  We demonstrate our technique on a transmission spectrum of the hot gas-giant exoplanet WASP-12b using a five-parameter model (temperature, a constant cloud opacity and the volume mixing ratios or relative abundance by number of water, ammonia and hydrogen cyanide) \cite{hk17}.  We obtain results consistent with the standard nested-sampling retrieval method.  Additionally, we can estimate the sensitivity of the measured spectrum to constraining the model parameters and we can quantify the information content of the spectrum.  Our method can be straightforwardly applied using more sophisticated atmospheric models and also to interpreting an ensemble of spectra without having to retrain the random forest.}

We use the previously analysed Hubble Space Telescope Wide Field Camera 3 (WFC3) transmission spectrum of the hot Jupiter WASP-12b, where the volume mixing ratio of water was inferred to be $\sim 10^{-4}$ to $\sim 10^{-2}$ and the temperature $\sim 1000$ K \cite{k15}.  Transmission spectra measure the wavelength-dependent obscuration of starlight by a transiting exoplanet, which encodes signatures of absorption by molecules and clouds in the exoplanetary atmosphere.  The choice of this spectrum was to ensure continuity between previous studies \cite{hk17,k15} and because we expect WFC3 to be the workhorse for measuring exo-atmospheric spectra for the immediate future.  We implement the random forest method \cite{ho98,breiman01}, which is a supervised form of machine learning. It combines the use of a decision tree \cite{bf84} and bootstrapping with replacement, and may be used on both discrete and continuous training sets.  A decision tree is a way of splitting a training set into subsets based on common characteristics of its members \cite{kelle15}.  The splitting is performed so as to maximize the gain in information entropy \cite{kelle15}.  Since decision trees are sensitive to slight changes in the training set, they are suitable for use with the bootstrapping method, which constructs the decision tree by randomly drawing from the training set \cite{kelle15}.

The training set consists of 80,000 synthetic WFC3 transmission spectra, each described by 5 parameters: the temperature ($T$), volume mixing ratios (relative abundances by number) of water ($X_{\rm H_2O}$), ammonia ($X_{\rm NH_3}$) and hydrogen cyanide ($X_{\rm HCN}$), and a constant cloud opacity ($\kappa_0$).  Given that these 5 parameters represent continuous data, we make use of ``regression trees" rather than decision trees (which are used for discrete data) in our random forest \cite{kelle15}. For each spectrum, the values of the 5 parameters are randomly generated either from a log-uniform (volume mixing ratios and cloud opacities) or uniform (temperature) distribution.  In addition to adopting the same wavelength range and 13 bins of the measured WASP-12b WFC spectrum \cite{k15}, we assume a noise floor of 50 parts per million (ppm) on the transit depth.

In a general machine-learning situation, each member of a training set is associated with a number of characteristics known as ``features" (in the jargon of machine learning), e.g., color, height, type of terrain.  For a spectrum, the features are the number of data points it contains.  Here, the WFC3 spectrum has 13 features or binned data points.  Within the training set, each synthetic spectrum is identified by its values of the 5 parameters.  The training set of 80,000 synthetic spectra resides in a 13-dimensional space, where each dimension corresponds to a wavelength bin.  Along the axis of each dimension is a continuous range of values of the transit radii.  The goal is to relate an entry (a synthetic spectrum) in this 13-dimensional space with the range of values of each of the 5 parameters.  We accomplish this by sub-dividing the 13-dimensional space into patches or islands, which is handled using a regression tree.  Each patch encompasses some subset of the training set, from which the variance in their transit radii may be computed.  The sub-division of the 13-dimensional space is done so as to minimize the sum of the variances of all of these patches.  This is conceptually equivalent to maximizing the gain in information entropy for discrete data \cite{kelle15}.

Upon setting up the regression tree, we use it in tandem with a bootstrapping method.  To train each regression tree, we randomly draw from the 80,000 synthetic spectra in the training set.  Upon each draw, the drawn synthetic spectrum is placed back into the training set, allowing for it to be drawn more than once.  Each regression tree may be visualized as being a predictive ``voter", who returns the ranges of parameter values given the 13 data points of the measured WFC3 transmission spectrum.  While a single regression tree produces predictions with large uncertainties, random forests mitigate this pitfall by combining the responses of multiple trees. We performed tests that indicate a convergence of these predictions using 1000 regression trees (see Methods).  Using 1000 regression trees to form a random forest, we are able to compute the posterior distributions of the parameters \cite{c11}.

\begin{figure*}
%\vspace{-0.1in}
\begin{center}
\includegraphics[width=1.8\columnwidth]{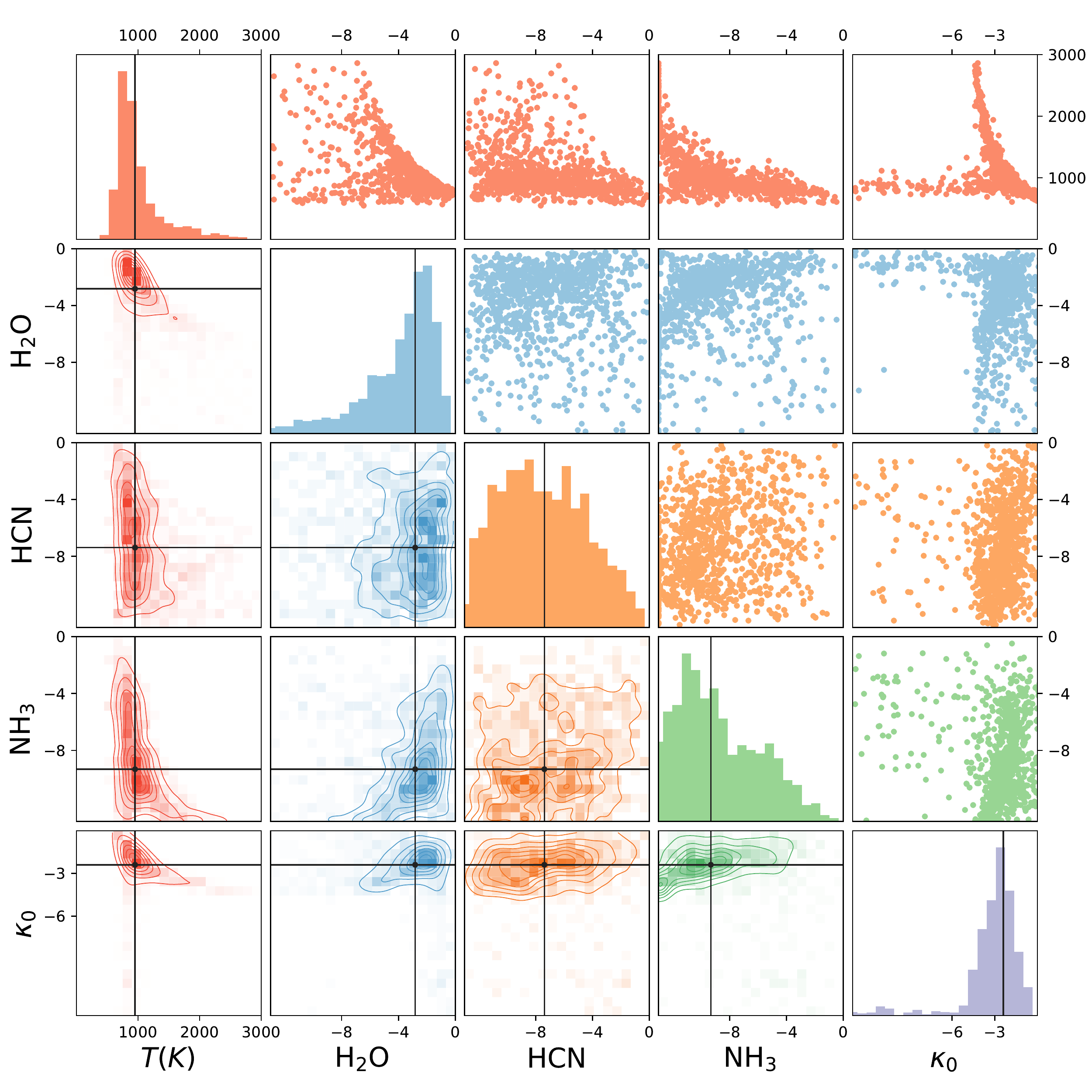}
\end{center}
\vspace{-0.1in}
\caption{Posterior distributions of the relative molecular abundances (volume mixing ratios), temperature and cloud opacity obtained from the machine-learning retrieval analysis of the WFC3 transmission spectrum of WASP-12b. Shown are the logarithm (base 10) of the volume mixing ratios and cloud opacity.  Within each scatter plot, each dot is an individual prediction of a single regression tree in the random forest.  The straight lines indicate the median values of the parameters.  Note that the volume mixing ratios and cloud opacity are associated with a factor $(P_0/10 \mbox{ bar})$ due to the normalization degeneracy.}
\label{fig:posteriors}
\end{figure*}

Figure \ref{fig:posteriors} shows the posterior distributions of the temperature, cloud opacity and volume mixing ratios of water, ammonia and hydrogen cyanide.  The retrieved water volume mixing ratio ($\log{X_{\rm H_2O}}=-2.8^{+1.4}_{-3.6}$) and temperature ($T=952^{+412}_{-151}$ K) values are broadly consistent with the previous analysis \cite{k15}.  A non-zero cloud opacity ($\log{\kappa_0} =-2.3^{+1.1}_{-1.6}$) is necessary to flatten the spectral continuum blueward of the 1.4 $\mu$m water feature.  The degeneracies between the temperature, molecular abundances and cloud opacity are consistent with physical intuition.  As the temperature increases linearly, the molecular opacities increase exponentially, a property that may be compensated by an order-of-magnitude decrease in the volume mixing ratio.  An increasing temperature also reduces the differences in opacity between the water feature and the spectral continuum blueward of it, a property that may be mimicked to some extent by the cloud opacity.  Clouds blunt the strength of molecular features, which may be compensated by order-of-magnitude increases in the abundances.

The retrieved volume mixing ratios of ammonia and hydrogen cyanide are several orders of magnitude lower than that of water: $\log{X_{\rm HCN}}=-7.6^{+3.3}_{-3.0}$, $\log{X_{\rm NH_3}}=-9.2^{+4.2}_{-2.9}$.  Running a pair of nested-sampling retrievals shows that the Bayes factor \cite{trotta08} between a model with water only versus one with all three molecules is 0.6 (with the former having the higher Bayesian evidence), implying that there is a lack of evidence for strongly favouring one model over the other.  Essentially, there is no evidence for claiming the detection of either hydrogen cyanide or ammonia.

\begin{figure*}
%\vspace{-0.1in}
\begin{center}
\includegraphics[width=1.8\columnwidth]{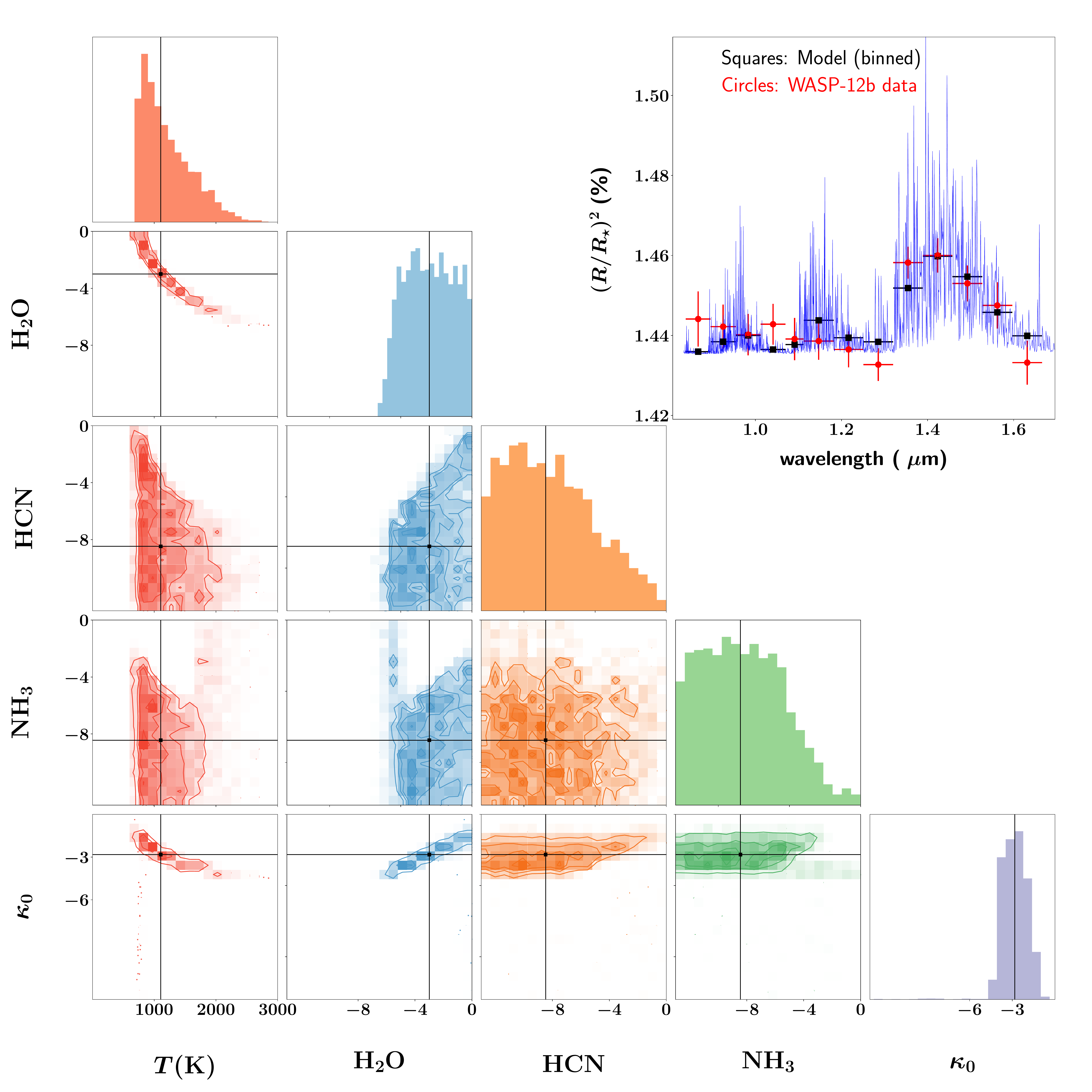}
\end{center}
\vspace{-0.1in}
\caption{Same as Figure \ref{fig:posteriors}, but for nested-sampling retrieval.  Additionally, the insert shows the measured versus best-fit model transmission spectra.}
\label{fig:heliost}
\end{figure*}

As a consistency check, Figure \ref{fig:heliost} shows the posterior distributions of parameters from our nested-sampling retrieval \cite{skilling06,feroz08}.  The retrieved parameter values from the nested-sampling retrieval are $T=1105^{+545}_{-287}$ K, $\log{X_{\rm H_2O}}=-3.0^{+2.0}_{-1.9}$, $\log{X_{\rm HCN}}=-8.5^{+3.8}_{-2.9}$, $\log{X_{\rm NH_3}}=-8.4^{+3.1}_{-2.9}$, $\log{\kappa_0} =-2.8 \pm 0.9$.  It is worth noting that the interpretation of transmission spectra suffers from a ``normalization degeneracy" \cite{bs12,hk17}.  To break this normalization degeneracy requires that one specifies a unique relationship between a reference transit radius ($R_0$) and reference pressure ($P_0$), which cannot be directly inferred from the WFC3 data alone.  In practice, what this means is that instead of the volume mixing ratio of molecules ($X_i$), one retrieves the quantity $X_i P_0$.  In the results shown, we have set $R_0 = 1.79 ~R_{\rm J}$ and $P_0=10$ bar to facilitate comparison with a previous study \cite{k15}.

Having demonstrated that we can use supervised machine learning to perform atmospheric retrieval, we now push beyond the regular analysis.  First, we would like to check the values of the 5 parameters predicted by the random forest method versus ``ground truth" values.  For the latter, we generate another 20,000 WFC3 synthetic transmission spectra.  We then apply our random forest method, previously trained on the 80,000 synthetic spectra, to predict the parameter values of these 20,000 new synthetic spectra.  Figure \ref{fig:sensitivity} shows that there is a one-to-one correspondence between the predicted and real values, albeit with some scatter.  To verify that the scatter is due to intrinsic model degeneracies (physics) and not due to our implementation of the random forest method itself, we performed other suites of calculations with different numbers of regression trees and noise floors (see Methods).

This comparison between the predicted versus real parameter values provides a rough estimate of the minimum values of the parameters that the retrieval is sensitive to, given the noise model assumed (a constant 50 ppm in our case).  For example, the linear trend between the predicted versus real values of the volume mixing ratios of water, hydrogen cyanide and ammonia starts to flatten below $\sim 10^{-6}$, suggesting that volume mixing ratios below one part in a million are undetectable given the WFC3 transmission spectrum of WASP-12b.

\begin{figure*}
%\vspace{-0.1in}
\begin{center}
\includegraphics[width=1.8\columnwidth]{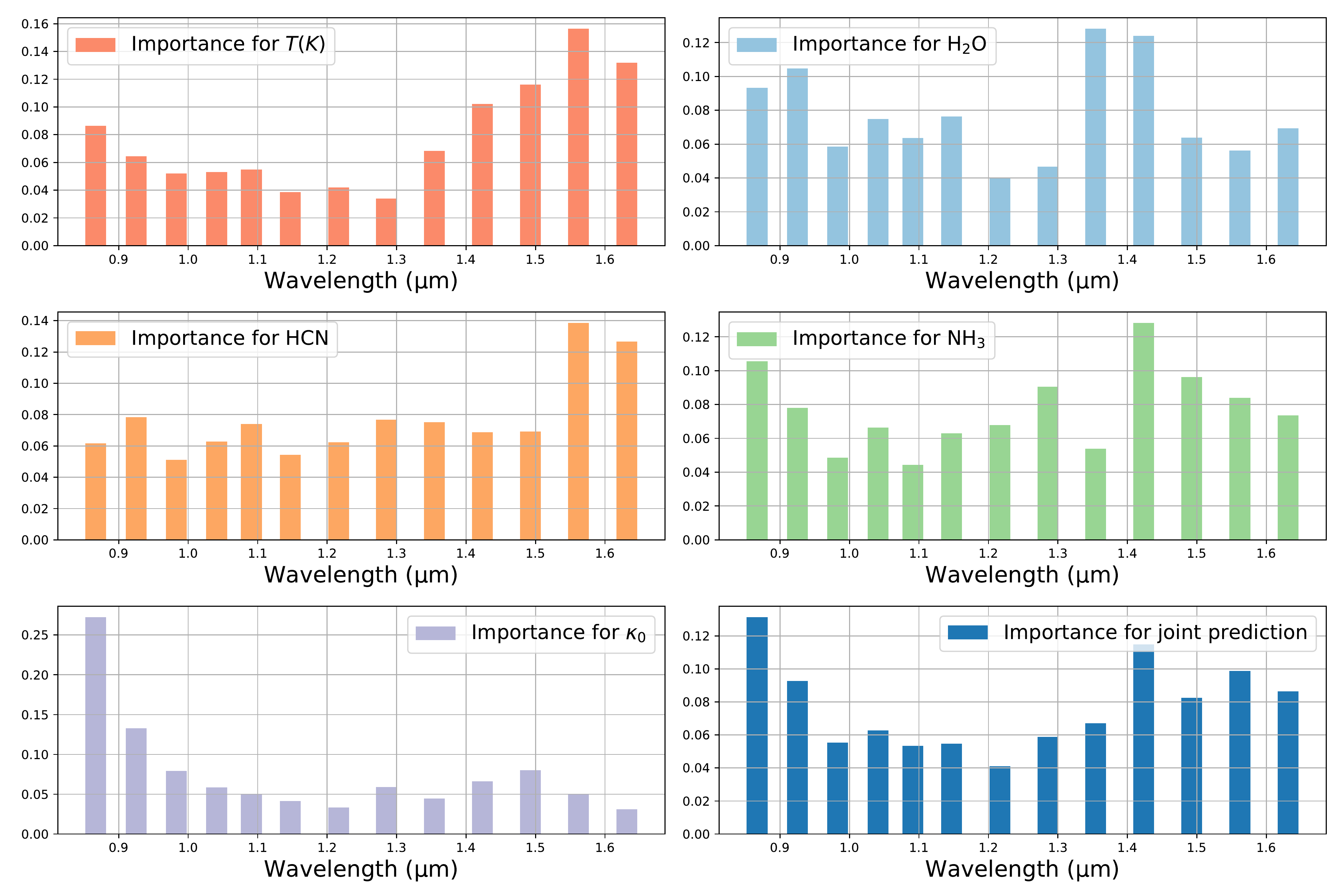}
\end{center}
\vspace{-0.1in}
\caption{Feature importance plots associated with the machine-learning retrieval analysis of the WFC3 transmission spectrum of WASP-12b. Values along the vertical axis indicate the relative importance of a data point for retrieving the value of a given parameter. Within each panel, the vertical axis values sum up to unity.}
\label{fig:infocontent}
\end{figure*}

Second, we can use our approach to analyse the information content of the measured WFC3 transmission spectrum.  While information content analysis has been previously considered \cite{line12,bl17,howe17}, we offer a complementary analysis and show that this is a natural outcome of the random forest method, which is called the ``feature importance" analysis.  Figure \ref{fig:infocontent} shows the relative weight of each of the 13 data points in the WFC3 transmission spectrum towards determining the value of each parameter.  Physical intuition tells us that the data points at around 1.4\,$\mu$m are the most constraining for the water abundance.  The feature importance analysis shows that the two data points near 1.4 $\mu$m contain about 30\% of the information that goes towards constraining the volume mixing ratio of water.  The two bluest data points contain more than 40\% of the information needed to constrain the cloud opacity, because they quantify the flatness of the spectral continuum.  The two reddest data points are most constraining for hydrogen cyanide.

There are straightforward extensions of random-forest retrieval for which no conceptual obstacles exist.  We have demonstrated the method on a spectrum with 13 data points, but the random forest method has been shown to work well even for 1000--10,000 data points \cite{hastie01,s13,z14,r15,zhang17}.  This property implies that random-forest retrieval is applicable to future James Webb Space Telescope (JWST) spectra spanning a broader range of wavelengths with $\sim 100$--1000 data points \cite{greene16}.  The information content analysis may be used to influence observational campaigns and the design of spectrographs, depending on the intended scientific goal.

Another straightforward extension is to train a random forest once and apply it to an ensemble of spectra.  In the current study, we picked a specific object (WASP-12b) to demonstrate our method.  There is no conceptual obstacle to making model grids where the surface gravity is allowed to vary.  The random forest is trained on this larger grid, but the value of the surface gravity may be fixed to the measured value of a specific object during analysis with no need for retraining.  In the study of stars and brown dwarfs, model grids spanning different ages, luminosities, radii, gravities and cloud configurations have traditionally been used to analyse ensembles of objects \cite{marley96,burrows97,baraffe02}.  It is conceivable that one may use model grids produced by different research groups to perform retrievals, even if the computer codes used to generate these grids are proprietary.

For the current study, we have showed that more sophisticated models are not necessary to analyse the WFC3 spectrum of WASP-12b.  However, there is nothing that prevents one from considering more sophisticated models.  For example, using the non-isothermal model of \cite{hk17} in tandem with the non-grey cloud model of \cite{kh18} would add 4 more parameters to the retrieval.  A longstanding shortcoming of atmospheric retrieval, which is the non-self-consistency of the physics and chemistry in the models, may now be obviated using random-forest retrieval.

\vspace{0.2in}

\noindent
{\scriptsize \textbf{Data Availability:} The data that support the plots within this paper and other findings of this study are available from the corresponding author upon reasonable request. 

\vspace{0.2in}

\noindent
\textbf{Code Availability:} The code used to generate all random-forest retrievals can be accessed at \texttt{https://github.com/exoclime}.

\vspace{0.2in}

\noindent
Correspondence and request for materials should be made to K.H.  We acknowledge partial financial support from the Center for Space and Habitability (P.M.N. and K.H.), the University of Bern International 2021 Ph.D Fellowship (C.F.), the PlanetS National Center of Competence in Research (K.H.), the Swiss National Science Foundation (R.S., C.F. and K.H.), the European Research Council via a Consolidator Grant (K.H.) and the Swiss-based MERAC Foundation (K.H.).}

\vspace{0.2in}

\noindent
{\scriptsize P.M.N. led the development of computer codes used for this study, performed the machine-learning related calculations, participated in the experimental design and made the majority of the figures.  C.F. computed the grid of atmospheric models used as the training set, participated in the experimental design and performed the nested-sampling retrievals. R.S. co-led the scientific vision and experimental design, and co-wrote the manuscript.  K.H. co-led the scientific vision and experimental design and led the writing and typesetting of the manuscript.}

\section*{Methods}

\begin{table*}%[!h]
\label{tab:tests}
\begin{center}
\caption{Resolution test for spectral resolution of opacities}
\begin{tabular}{lccccc}
\hline
\hline
Resolution & $T$ (K) & $\log{X_{\rm H_2O}}$ & $\log{X_{\rm HCN}}$ & $\log{X_{\rm NH_3}}$ & $\log{\kappa_0}$  \\
\hline
\hline
1 cm$^{-1}$ & $1075^{+427}_{-259}$ & $-2.8^{+1.9}_{-1.7}$ & $-8.0^{+3.7}_{-3.3}$ & $-8.9^{+3.3}_{-2.7}$ & $-2.8 \pm 0.8$ \\
2 cm$^{-1}$ & $1114^{+413}_{-269}$ & $-3.0^{+1.8}_{-1.6}$ & $-8.1^{+3.8}_{-3.2}$ & $-8.7^{+3.3}_{-2.8}$ & $-2.9^{+0.8}_{-0.7}$ \\
5 cm$^{-1}$ & $1105^{+545}_{-287}$ & $-3.0^{+2.0}_{-1.9}$ & $-8.5^{+3.8}_{-2.9}$ & $-8.4^{+3.1}_{-2.9}$ & $-2.8\pm 0.9$ \\
10 cm$^{-1}$ & $1228^{+548}_{-355}$ & $-3.5^{+2.2}_{-1.7}$ & $-8.5^{+3.8}_{-2.8}$ & $-8.8^{+3.2}_{-3.0}$ & $-3.2^{+0.8}_{-0.7}$ \\
\hline
\hline
\end{tabular}\\
%\vspace{0.05in}
\end{center}
\end{table*}

For the physics input, we choose to use a previously validated analytical formula to convert the temperatures, molecular opacities and relative abundances of molecules into transit radii \cite{hk17}.  The simplicity of this forward model allows us to straightforwardly diagnose problems and understand trends in the posterior distributions.  We use the simplest incarnation of this formula, which assumes that the atmosphere is isothermal, isobaric and hosts a grey cloud.  Using the nested sampling method \cite{skilling06,feroz08,feroz09,b14}, we have performed regular retrievals, which indicate that non-isothermal behavior and non-grey clouds are not necessary to explain the data given its current level of quality and sophistication.  We include the opacities of water (H$_2$O), hydrogen cyanide (HCN) and ammonia (NH$_3$), computed using the \texttt{ExoMol} spectroscopic line lists \cite{barber06,barber14,y11,y13,yt14} as input and in the standard way, meaning that the opacities are products of the integrated line strength and line shape, and the line shapes are assumed to be truncated Voigt profiles \cite{rothman98,gh15}.

For each model, we randomly pick values of the parameters over the following ranges: $T= 500$--2900 K, $X_{\rm H_2O}=10^{-13}$--$1$, $X_{\rm HCN}=10^{-13}$--$1$, $X_{\rm NH_3}=10^{-13}$--$1$, $\kappa_0 = 10^{-13}$--$10^2$ cm$^2$ g$^{-1}$.  The surface gravity of WASP-12b is taken to be 977 cm s$^{-2}$ \cite{hebb09}.  The spectroscopic database used to construct the NH$_3$ opacities does not exist for temperatures above 1600 K \cite{y11}.  For computational reasons, we set the NH$_3$ opacity to be zero and the volume mixing ratio to be small ($10^{-13}$) if the temperature exceeds this threshold.  Fortunately, ammonia is expected to be a minor species at high temperatures, where the dominant nitrogen carrier is instead expected to be molecular nitrogen \cite{bs99,ht16}.

The ``features" are the 13 values of the transit radius, across wavelength, associated with each transmission spectrum.  One may visualize 13 columns, each with 80,000 values of the transit radius.  One then visualizes a 13-dimensional space, where each dimension is marked by a set of numerical thresholds.  Boundaries in this 13-dimensional space are drawn based on splitting the training set in order to minimize the total variance.  Each time a boundary is drawn, one is splitting the regression tree.  Once the reduction in the variance of the tree node is negligibly small (0.01 in our case), we stop splitting the training set.  Tree pruning methods are not used.  Each time the tree is split, only a random subset (4, which is about $\sqrt{13}$) of the 13 spectral bins is used.  In other words, both the members of the 80,000 training set, as well as the subset of spectral bins associated with each member, are randomly drawn in order to decrease the correlations between predictions from different trees.  For a pedagogical summary of the random forest method, please see \cite{kelle15}.  The implementations of the random forest method and $R^2$ metric are from the open-source \texttt{scikit.learn} library in the \texttt{Python} programming language.

It has been previously shown that the random forest method is capable of handling systems with 1000--10,000 features and tree depths of several tens to hundreds \cite{hastie01,s13,z14,r15,zhang17}.  Our current problem has 13 features and the regression trees have, on average, about 19,000 nodes and depths of 14.  

To check the robustness of our results with respect to our implementation of the random forest method, we examine retrieval outcomes with different numbers of regression trees.  Like before, we train on 80,000 synthetic spectra and then use it to analyse 20,000 more synthetic spectra.  Figure \ref{fig:trees} shows that the outcomes of these mock retrievals converge when the number of trees used exceeds about 100.  In the same figure, we also checked the retrieval outcomes with different levels of assumed noise floors.  For each of the 13 data points in the synthetic WFC3 spectra, we assume a Gaussian uncertainty on the transit depths with full widths at half-maximum of 10, 50 and 100 ppm, which represent ideal, typical and easily attainable conditions.  As expected, the variance associated with the true versus predicted values of the five parameters decreases (i.e., the coefficient of determination increases) when the assumed noise floor is lower.  As a further check, we first train a random forest on a model grid with an assumed noise floor of 50 ppm and use it to analyse mock data with assumed noise floors of 10, 50 and 100 ppm.  The resulting $R^2$ values are 0.676, 0.651 and 0.586, respectively, for the joint predictions.

\begin{figure*}%
%\vspace{-0.1in}
\begin{center}
\includegraphics[width=1.8\columnwidth]{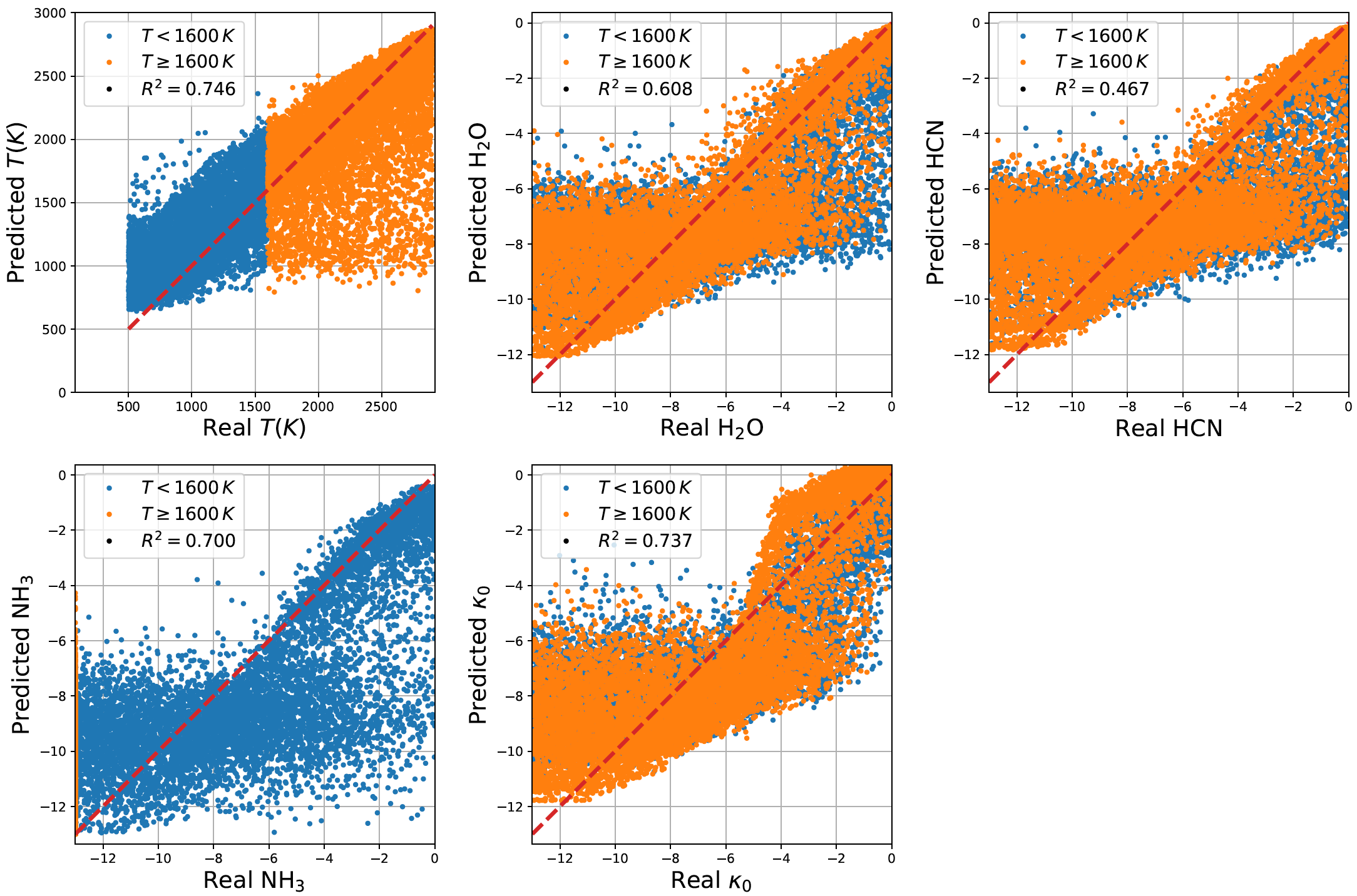}
\end{center}
\vspace{-0.1in}
\caption{True versus random-forest predicted values of the five parameters in our transmission spectrum model.  The coefficient of determination ($R^2$) varies from 0 to 1, where values near unity indicate strong correlations between the predicted and real values of a given parameter, based on the variance of outcomes.}
\label{fig:sensitivity}
\end{figure*}

\begin{figure*}
%\vspace{-0.1in}
\begin{center}
\includegraphics[width=\columnwidth]{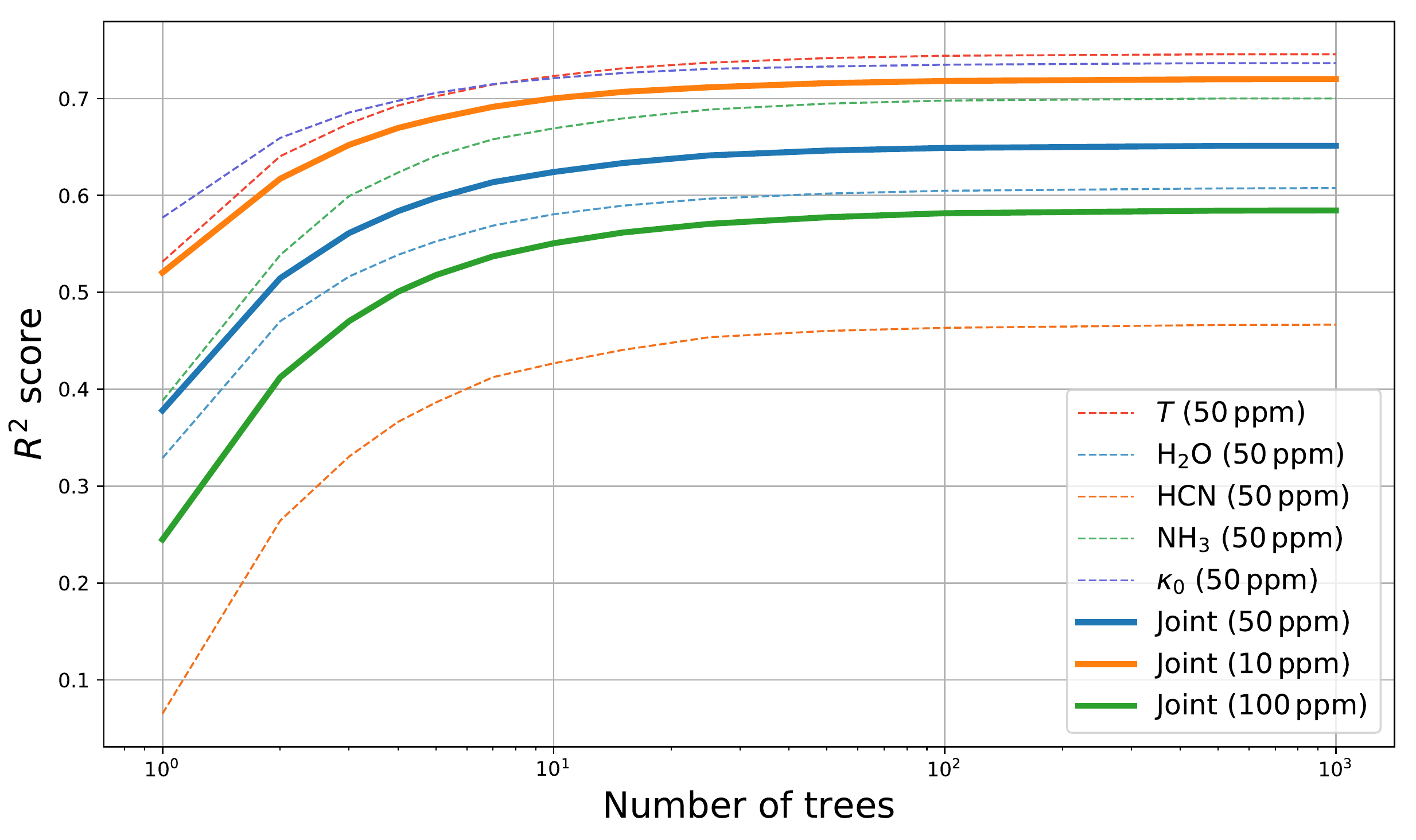}
\includegraphics[width=\columnwidth]{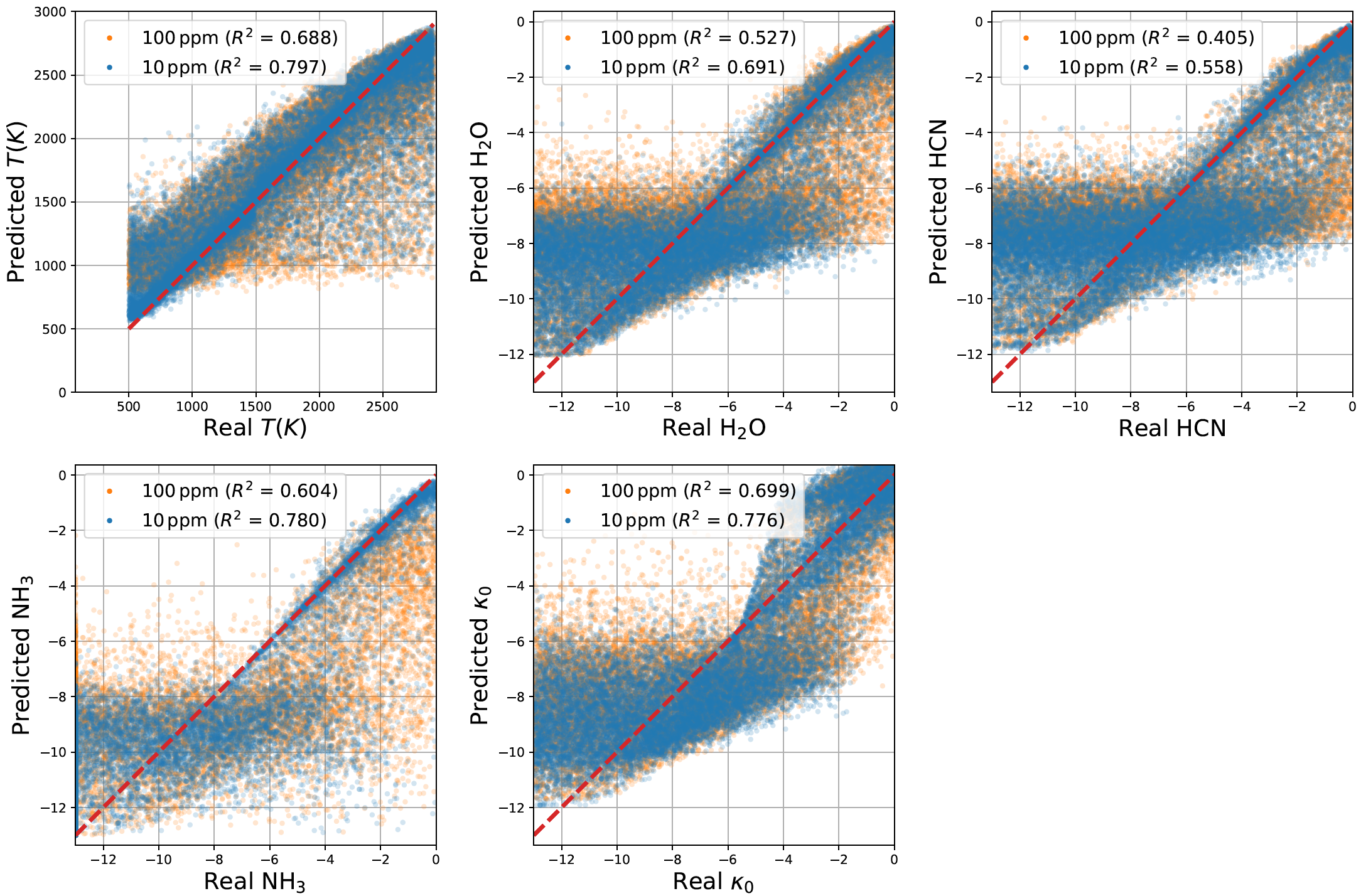}
\end{center}
\vspace{-0.1in}
\caption{Left panel: Coefficient of determination for each of the 5 parameters, as well as for the joint prediction, versus the number of regression trees used in the random forest with an assumed noise floor of 50 ppm.  Also included are the joint predictions for assumed noise floors of 10 and 100 ppm.  Right panel: Same as Figure \ref{fig:sensitivity}, but comparing mock retrievals with assumed noise floors of 10 versus 100 ppm.  The coefficient of determination ($R^2$) varies from 0 to 1, where values near unity indicate strong correlations between the predicted and real values of a given parameter.}
\label{fig:trees}
\end{figure*}

\begin{figure*}
%\vspace{-0.1in}
\begin{center}
\includegraphics[width=\columnwidth]{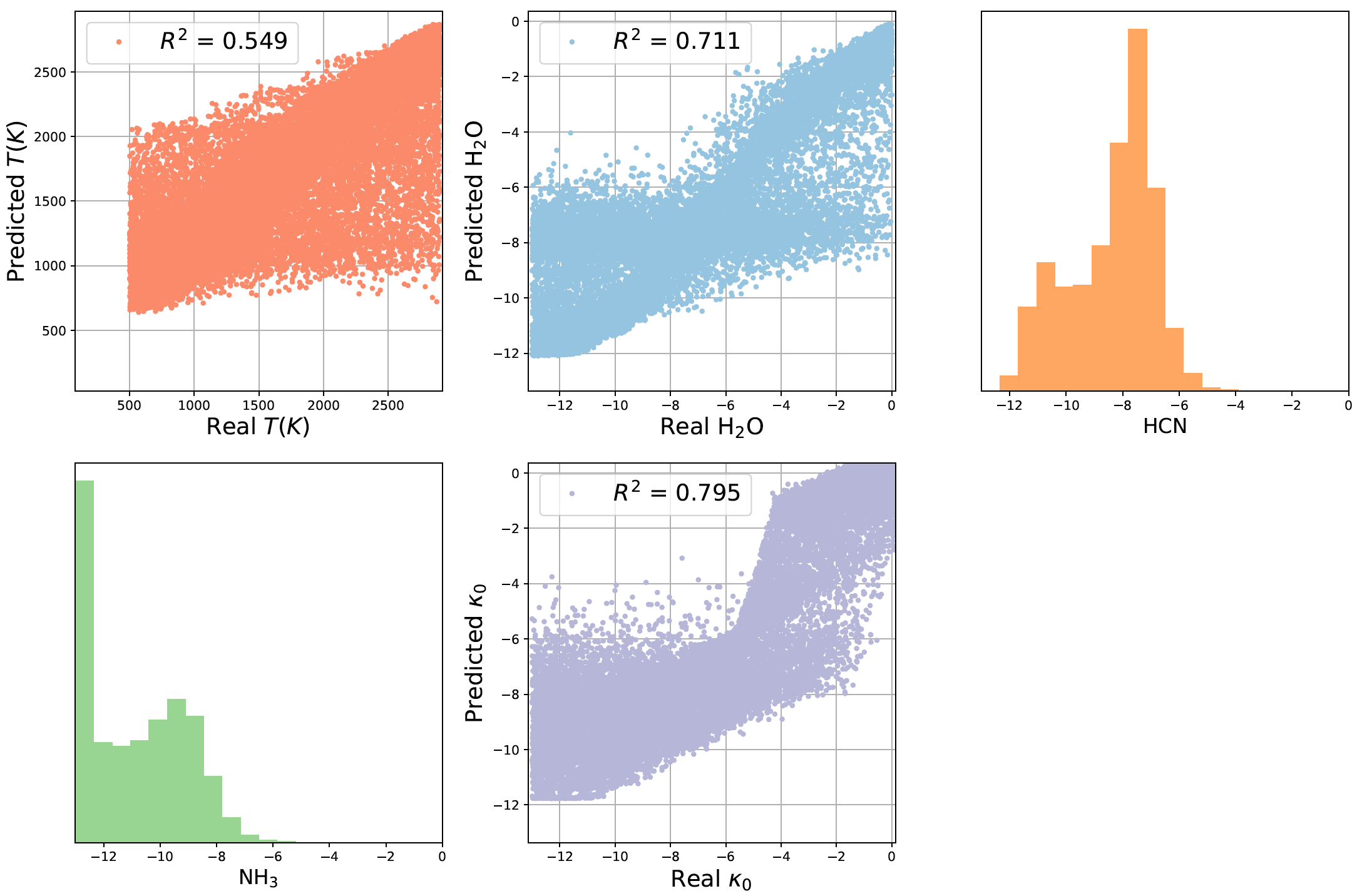}
\includegraphics[width=\columnwidth]{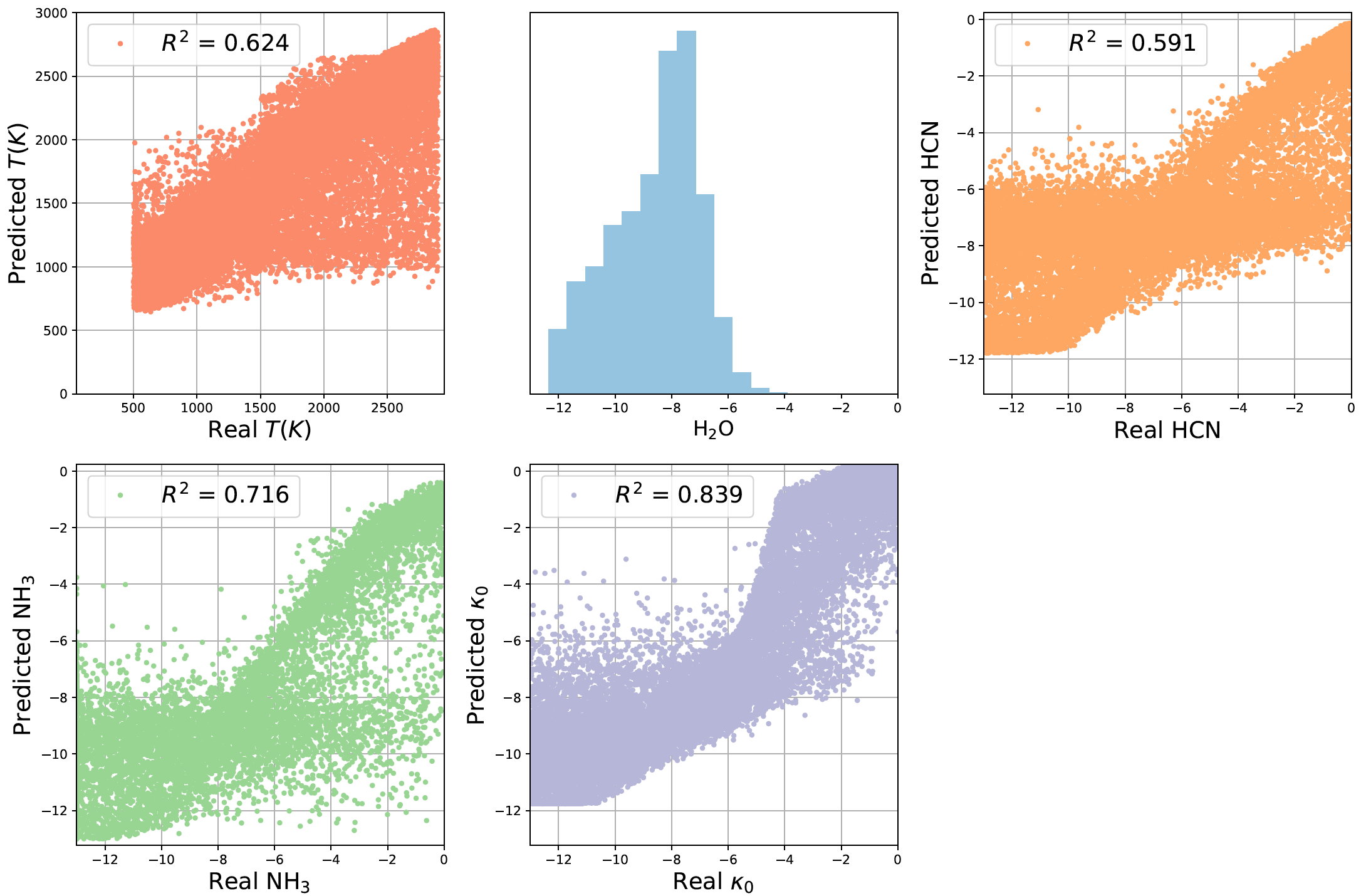}
\end{center}
\vspace{-0.1in}
\caption{True versus random-forest predicted values of the five parameters in our transmission spectrum model.  Left montage: H$_2$O only, where the posterior distributions of HCN and NH$_3$ are shown.  Right montage: HCN and NH$_3$ only, where the posterior distribution of H$_2$O is shown.}
\label{fig:tests}
\end{figure*}

We also ran the same mock retrievals for model grids where the atmosphere contains water only versus one that contains hydrogen cyanide and ammonia (without water), as shown in Figure \ref{fig:tests}.  In the former case, the retrievals return $X_{\rm HCN} \sim 10^{-8}$ and $X_{\rm NH_3} \sim 10^{-10}$ even when neither molecule is present in the mock spectra, which is consistent with our finding in Figure \ref{fig:sensitivity} that volume mixing ratios below $\sim 10^{-6}$ indicate non-detections of these molecules.  In the latter case, we obtain $X_{\rm H_2O} \sim 10^{-8}$, which is consistent with the non-detection of water.

As a final test and precursor for future studies, we generated mock JWST-like data in the NIRSpec range of wavelengths (0.8 to 5.0 $\mu$m) at a resolution of 100 (not shown).  Despite the increase in the number of features (data points) from 13 to 181, the time needed to train the random forest on 80,000 mock spectra only increased by a factor of 4 (without any attempt to parallelise the computation).  The time needed for interpreting the additional 20,000 mock spectra (termed ``testing") is virtually the same in both cases.  Furthermore, we note that both the training and testing steps are highly parallelisable.

\begin{figure*}
%\vspace{-0.1in}
\begin{center}
\includegraphics[width=1.8\columnwidth]{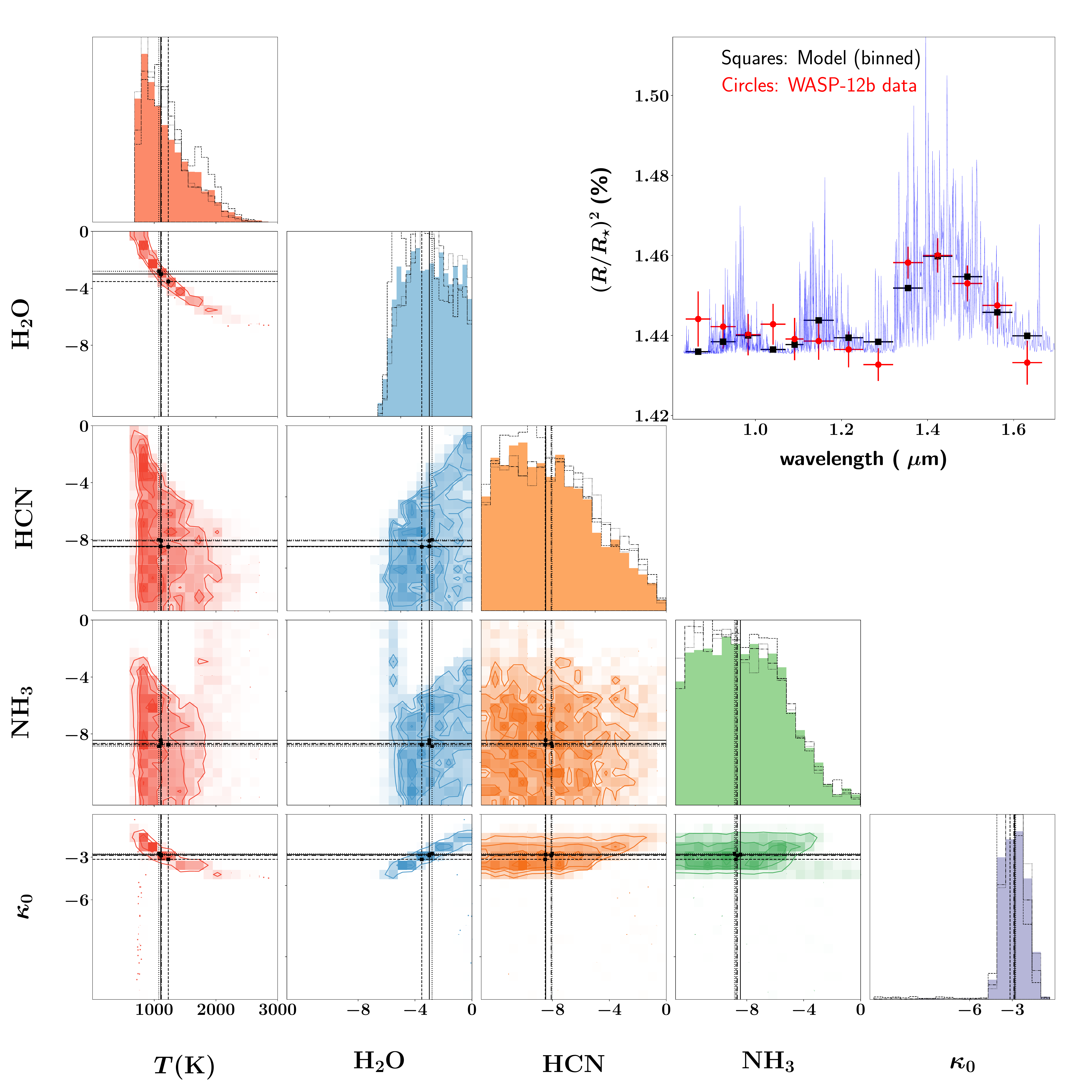}
\end{center}
\vspace{-0.1in}
\caption{Nested-sampling retrievals using opacities with spectral resolutions of 1 cm$^{-1}$ (dotted curves), 2 cm$^{-1}$ (dot-dashed curves), 5 cm$^{-1}$ (filled posteriors in colour and solid lines for the median values of parameters) and 10 cm$^{-1}$ (dashed curves), respectively.}
\label{fig:restest}
\end{figure*}

To determine the spectral resolution used for our opacities, we ran retrievals with resolutions of 1, 2, 5 and 10 cm$^{-1}$ assuming an isothermal atmosphere containing grey clouds and all three molecules.  Retrieval practitioners typically use a spectral resolution of 1 cm$^{-1}$ for their opacities \cite{line13,waldmann15,lavie17}, although it is not uncommon for workers to not state the spectral resolution used.  For these 4 resolutions, the retrievals are shown in Figure \ref{fig:restest}.  The corresponding retrieved parameter values are tabulated.  Based on this resolution test, we adopt 5 cm$^{-1}$ as our spectral resolution for the opacities. 

We assume pressure broadening to be negligible.  Since the inferred atmospheric temperature does not fall well below 1000 K and the volume mixing ratios are typically much smaller than unity, this is not an unreasonable assumption \cite{hk17}.  Operationally, to implement this assumption we assume a pressure of 1 mbar when computing the opacities.  As is accepted practice \cite{sb07}, our ignorance of the physics of pressure broadening forces us to truncate the Voigt profile at some distance from line center.  We have made an ad hoc choice of 100 cm$^{-1}$, but since pressure broadening is assumed to be negligible this has little to no effect on the outcome.

To check our assumption of a constant/grey cloud opacity, we ran another retrieval calculation with the non-grey cloud model of \cite{kh18}.  The Bayes factor for the pair of models with grey versus non-grey clouds is 0.6 (with the former having a higher Bayesian evidence), which implies there is no evidence for the data favouring the non-grey over the grey cloud model \cite{trotta08}.  In fact, we note that the model with non-grey clouds and water only has the same Bayesian evidence as one with grey clouds and all three molecules.  Similarly, the Bayes factor for a pair of models with isothermal versus non-isothermal atmospheres (both with water only) is 0.7, implying a lack of evidence for non-isothermal behavior.  The latter has lower Bayesian evidence and was computed using the non-isothermal analytical formula derived by \cite{hk17}.

%%% REFERENCES %%%

\end{document}